\documentclass[aps,prc,reprint,showkeys,showpacs,nofootinbib,superscriptaddress,floatfix]{revtex4-1}

\usepackage[english]{babel}
\usepackage{amsmath,amssymb,amsfonts}
\usepackage{bm}
\usepackage{graphicx}
\usepackage{braket}
\usepackage[colorlinks=true,allcolors=blue]{hyperref}
\usepackage{microtype}

\newcommand{\der}[1]{\mathrm{d}#1\;}
\newcommand{\be}{\begin{equation}}
\newcommand{\ee}{\end{equation}}
\newcommand{\Tr}{\operatorname{Tr}}

\begin{document}

\title{ Statistical theory of deformation distributions in nuclear spectra}
\date{\today}
\author{M.\ T.\ Mustonen} \email{mika.mustonen@yale.edu}
\affiliation{Center for Theoretical Physics, Sloane Physics Laboratory,  Yale University, New Haven, Connecticut 06520, USA}
\author{C.\ N.\ Gilbreth} \email{gilbreth@uw.edu}
\affiliation{Institute for Nuclear Theory, Box 351550, University of Washington, Seattle, Washington 98195, USA}
\author{Y.\ Alhassid} \email{yoram.alhassid@yale.edu}
\affiliation{Center for Theoretical Physics,  Sloane Physics Laboratory,  Yale University, New Haven, Connecticut 06520, USA}
\author{G.\ F.\ Bertsch} \email{bertsch@uw.edu}
\affiliation{Department of Physics and Institute for Nuclear Theory, Box 351560, University of Washington, Seattle, Washington 98195, USA}

\begin{abstract}
The dependence of the nuclear level density on intrinsic deformation is an important input to dynamical nuclear processes such as fission. Auxiliary-field Monte Carlo (AFMC) method is a powerful method for computing nuclear level densities. However, the statistical distribution of intrinsic shapes
is not readily accessible due to the formulation of AFMC in a spherical configuration-interaction shell-model approach.  Instead, theory of deformation up to now has largely relied on a mean-field approximation which breaks rotational symmetry. We show here how the distributions of the intrinsic quadrupole deformation parameters can be calculated within the AFMC method, and present results for a chain of even-mass samarium nuclei ($^{148}$Sm, $^{150}$Sm, $^{152}$Sm, $^{154}$Sm) which includes spherical, transitional, and strongly deformed isotopes. The method relies on a Landau-like expansion of the Helmholtz free energy in invariant polynomials of the quadrupole tensor.  We find that an expansion to fourth order provides an excellent description of the AFMC results. 
\end{abstract}
\maketitle
 
\section{Introduction}

Nuclear level densities are an essential ingredient in the Hauser-Feshbach theory~\cite{Hauser1952} of statistical nuclear reactions. In particular, models of fission require the knowledge of the level density as a function of nuclear deformation.  

The auxiliary-field Monte Carlo (AFMC) method, also known in nuclear physics as shell-model Monte Carlo (SMMC)~\cite{Lang1993,Alhassid1994,Koonin1997,Alhassid2001,Alhassid2017}, is a powerful technique for microscopic calculations of nuclear level densities within the configuration-interaction (CI) shell model approach~\cite{Nakada1997,Alhassid1999}. 
The method has been applied to nuclei as heavy as the lanthanides~\cite{Alhassid2008,Ozen2013}.

Deformation is usually introduced in a  mean-field approximation that breaks rotational invariance. It is thus a challenge to calculate deformation-dependent level densities in the CI shell model framework which preserves rotational invariance without invoking a mean-field approximation.

In Refs.~\cite{Alhassid2014,Gilbreth2018} the distribution of the axial quadrupole operator $\hat Q_{2 0}=  \sum_i \left[2 \hat z_i^2 - (\hat x_i^2+\hat y_i^2)\right]$ in the laboratory  frame was calculated using AFMC and shown to exhibit model-independent signatures of deformation. The use of quadrupole invariants~\cite{Kumar1972,Cline1986}, which in turn can be related to lab-frame moments of $\hat Q_{20}$ (up to fifth order in deformation), allowed the extraction of effective intrinsic deformation parameters $\beta,\gamma$. Quadrupole invariants have been used in the context of the CI shell model for lighter nuclei; see Refs.~\cite{Hady2016,Schmidt2017} and references cited therein. Here we introduce a novel method to calculate the complete intrinsic-frame quadrupole distribution using a Landau-like expansion of its logarithm. This enables us to compute the nuclear state density as a function of excitation energy $E_x$ and intrinsic deformation $\beta,\gamma$.

We demonstrate our approach for an isotopic chain of even-mass samarium nuclei, $^{148,150,152,154}$Sm. Signatures of the crossover from spherical to deformed nuclei in this isotopic chain have been observed in AFMC calculations~\cite{Ozen2013,Gilbreth2018}.  

This article is organized as follows. In Sec.~\ref{sec:labframe}, we briefly review the AFMC method and its application to calculate the distribution $P(q_{20})$ of the axial quadrupole operator  $\hat Q_{20}$ in the laboratory frame.  In Sec.~\ref{sec:intrinsic_frame}, we introduce a novel method to determine the quadrupole tensor distribution as a function of temperature in the intrinsic frame.  In Sec.~\ref{sec:state_densities}, we use the saddle-point approximation to convert this temperature-dependent intrinsic frame distribution to density of states $\rho(E_x,\beta,\gamma)$ as a function of the excitation energy $E_x$ and intrinsic deformation parameters $\beta,\gamma$. Finally, in Sec.~\ref{sec:conclusions} we summarize our method in a more general context.  The AFMC data files and the scripts used to generate the results of the work presented here are included in the Supplementary Material repository of this article.

\section{Quadrupole projection in the laboratory frame}\label{sec:labframe}

\subsection{The AFMC method}

We briefly review the AFMC method, emphasizing the elements that are essential for our current application. For a recent review of AFMC in nuclei, see Ref.~\cite{Alhassid2017}.

A nucleus at finite temperature $T$ and Hamiltonian $\hat H$ is described by the Gibbs ensemble $\exp (-\hat H/T)$, which can also be viewed as a propagator in imaginary time $\beta=1/T$.\footnote{Here we adopt natural units $k_\mathrm{B} = 1$, and use the circumflex to denote operators in the many-particle Fock space.} The AFMC method is based on the Hubbard-Stratonovich (HS) transformation~\cite{Hubbard1958}, in which the propagator $\exp (-\beta \hat H)$ is decomposed into a superposition of one-body propagators $\hat U_\sigma$ that describe non-interacting nucleons in external time-dependent auxiliary fields $\sigma$
\be\label{HS}
  e^{-\beta \hat H} = \int \mathcal D [\sigma] \; G_\sigma  \hat U_\sigma \;,
\ee
where $G_\sigma$ is a Gaussian weight.

Using Eq.~(\ref{HS}), the thermal expectation value of an observable $\hat O$ is given by
\be \label{expect}
  \langle \hat  O \rangle = \frac{\operatorname{Tr} (\hat  O 
e^{-\beta \hat H})}{\operatorname{Tr} e^{-\beta \hat H}} 
= \frac{\int \mathcal D [\sigma] \; G_\sigma \Tr (\hat O \hat U_\sigma)}{\int \mathcal D [\sigma] \; G_\sigma  \Tr \hat U_\sigma} \;.
\ee

In AFMC, the expectation value in \eqref{expect} is evaluated by Monte Carlo sampling of the auxiliary fields $\sigma$ according to the positive-definite weight function $W_\sigma = G_\sigma |\Tr(\hat U_\sigma)|$. We define the $W$-weighted average of a quantity $X_\sigma$ by
\be\label{ave_x}
\left \langle X_\sigma \right\rangle_W \equiv \frac {\int D[\sigma ] W_\sigma  X_{\sigma} \Phi_{\sigma}} { \int D[\sigma]  W_\sigma \Phi_{\sigma}} \;,
\ee
where $\Phi_\sigma\equiv \Tr \hat U_\sigma /\vert \Tr \hat U_\sigma \vert$ is the Monte Carlo sign function. The thermal expectation in (\ref{expect}) can then be written as 
\be\label{observ-W}
 \langle \hat O \rangle =  \left\langle { \Tr ( \hat O \hat U_\sigma) \over \Tr  \hat U_\sigma} \right\rangle_W  \;.
 \ee
Denoting the sampled auxiliary-field configurations by $\sigma_k$, the expectation value in (\ref{observ-W}) is estimated by
 \be
  \langle \hat  O \rangle \approx
 \frac{\sum_k  \; \langle \hat O \rangle_{\sigma_k} \Phi_{\sigma_k}}{\sum_k \; \Phi_{\sigma_k}} \;,
\ee
where $\langle \hat O \rangle_\sigma =  \Tr ( \hat O \hat U_\sigma) / \Tr  \hat U_\sigma$.

An essential feature of the AFMC is that the many-particle traces $\Tr$ can be reduced to expressions involving only matrix algebra in the single-particle space.
For example, the grand-canonical trace of the many-particle propagator $\hat U_\sigma$ in Fock space is given by
\be
\Tr  \hat U_{\sigma} = \det (1 +{\bf U}_\sigma) \;,
\ee
where ${\bf U}_\sigma$ is the matrix representation of $\hat U_\sigma$ in the single-particle space.

Since nuclei are finite-size systems, it is important to evaluate the traces in Eq.~(\ref{observ-W}) in the canonical ensemble, i.e., at fixed particle number. We use discrete Fourier transforms to project on fixed number of protons and neutrons~\cite{Ormand1994,Alhassid1999}.  

\subsection {$\hat Q_{20}$ projection}

The mass quadrupole tensor operator is defined by
\be
\hat Q_{2 \mu} = \sqrt{\frac{16 \pi}{5}} \int  \text{d}^3\mathbf{r}  \hat \rho({\bf r}) r^2 Y_{2 \mu}(\theta,\varphi) \;,
\ee
where  $\hat \rho({\bf r}) =\sum_i \delta({\bf r}_i -{\bf r})$ is the total single-particle density (including both protons and neutrons) at point ${\bf r}$. 

The lab-frame probability distribution for measuring the eigenvalue $q_{2 0}$ of the axial quadrupole operator $\hat Q_{20} = \sum_i [2\hat z_i^2 - (\hat x_i^2 + \hat y_i^2)]$ is defined by
\be\label{q-dist}
P(q_{20}) = \frac{1}{Z} \Tr[\delta (\hat Q_{20} - q_{20}) e^{-\beta \hat H}] \;,
\ee
where $Z = \Tr e^{-\beta \hat H}$ is the partition function. Expanding in a basis of many-particle eigenstates
\begin{equation} \label{qdist_expansion}
  P(q_{20}) = \frac{1}{Z} \sum_n \delta(q_{20} - q_n) \sum_m \langle q_n | e_m \rangle^2 e^{-\beta e_m} \;,
\end{equation}
where $q_n$ and $|q_n\rangle$ are the eigenvalues and eigenstates of the operator $\hat Q_{20}$, and $e_m$ and $|e_m\rangle$ are the eigenvalues and eigenstates of the Hamiltonian $\hat H$.
Since $\hat Q_{20}$ does not commute with the Hamiltonian, $\langle q_n | e_m \rangle \ne \delta_{n,m}$.  

In AFMC, we calculate (\ref{q-dist}) from
\be\label{q-dist-AFMC}
  P(q_{20}) = \frac{1}{\langle \Phi_\sigma \rangle_W} \left\langle \frac{\operatorname{Tr} [\delta(\hat Q_{20} - q_{20}) \hat U_\sigma]}{\operatorname{Tr} \hat U_\sigma} \Phi_\sigma \right\rangle_W\;,
\ee
where the $\delta$ function is represented by a Fourier transform.  In practice, we divide the range $q_{20} \in [-q_\textrm{max}, q_\textrm{max}]$ to $2M+1$ equal intervals and evaluate the quadrupole-projected trace using a discretized Fourier decomposition
\be\label{Fourier}
  \operatorname{Tr} [\delta(\hat Q_{20} - q_{20}) \hat U_\sigma]
  \approx \frac{1}{2q_\textrm{max}} \sum_{k = -M}^M e^{-i \varphi_k q_{20}} \operatorname{Tr} (e^{i \varphi_k \hat Q_{20}} \hat U_\sigma) \;,
\ee
where $\varphi_k = \pi k/q_\textrm{max}$.
To aid the otherwise slow thermalization and decorrelation of the moments $\langle \hat Q_{20}^n \rangle$ with the pure Metropolis sampling, we augment the generated field configurations by rotating them through a certain set of angles~\cite{Alhassid2014,Gilbreth2018}.

\section{Quadrupole distributions in the intrinsic frame}\label{sec:intrinsic_frame}

\subsection{Intrinsic variables}

For given values of the quadrupole tensor $q_{2\mu}$ in the laboratory frame,\footnote{The quadrupole operators commute in coordinate space but not in the truncated CI shell model space. However, the effect of their non-commutation is small and will be ignored in the following.} we define dimensionless quadrupole deformation parameters $\alpha_{2\mu}$ from the liquid drop model
\be\label{q-to-alpha}
q_{2\mu} = \frac{3}{\sqrt{5 \pi}} r_0^2 A^{5/3} \alpha_{2\mu}\;,
\ee
where $r_0=1.2$ fm and $A$ is the mass number of the nucleus. For each set $\alpha_{2\mu}$ we can define an intrinsic frame whose orientation is specified by the Euler angles $\Omega$ and in which the quadrupole deformation parameters $\tilde \alpha_{2\mu}$ are 
\be
{\tilde \alpha}_{21} = {\tilde \alpha}_{2\,-1} = 0, \;\;\; {\tilde \alpha}_{22} = {\tilde \alpha}_{2\,-2} = \text{real} \,.
\ee

The intrinsic quadrupole deformation variables $\tilde \alpha_{2 \mu}$  are parametrized by the usual coordinates $(\beta, \gamma)$ defined by~\footnote{Following established conventions, we denote both the inverse temperature and the axial deformation parameter by the same symbol $\beta$. The intended meaning should be clear from the context at each occurrence throughout this article.}
\be\label{metric}
   \tilde \alpha_{20} = \frac{1}{\sqrt{2}} \beta \sin \gamma \;;\;\;\; \tilde \alpha_{22} = \tilde \alpha_{2,-2} = \beta \cos \gamma \;.
\ee
The transformation from the lab-frame $\alpha_{2\mu}$ to the intrinsic variables $\beta,\gamma,\Omega$ is characterized  by the metric
 \be\label{metric}
\prod_\mu d {\alpha_{2\mu}}=  \frac{1}{2}\beta^4 |\sin (3\gamma)|  \, d \beta \, d \gamma \, d \Omega \;.
\ee

\subsection{Distribution of the quadrupole deformation in the intrinsic frame}

We denote the distribution of the quadrupole deformation tensor in the laboratory frame at temperature $T$ by $P(T,\alpha_{2\mu})$. This distribution is invariant under rotations and therefore depends only on the intrinsic variables $\beta,\gamma$, i.e., $P(T,\alpha_{2\mu}) = P(T,\beta,\gamma)$. 

Using the metric (\ref{metric}), and integrating over the spatial angles $\Omega$, the probability distribution in the intrinsic variables $\beta,\gamma$ is given by
\be\label{shape-dist}
4\pi^2 \beta^4 |\sin(3\gamma)|  P(T,\beta,\gamma) \;.
\ee

Quadrupole invariants can be constructed by taking products of the second-rank tensor $\alpha_{2\mu}$ that couple to total angular momentum zero. Up to fourth order, these invariants are given by
\begin{subequations}\label{eq:invariants-intrinsic}
\begin{equation}
  \alpha \cdot \alpha  = \beta^2 \;,
\end{equation}
\begin{equation}
  [ \alpha \times \alpha]_2 \cdot \alpha = -\sqrt{\frac{2}{7}} \beta^3 \cos (3\gamma) \;,
\end{equation}
\begin{equation}
  (\alpha \cdot \alpha)^2 = \beta^4 \;.
\end{equation}
\end{subequations}
We note that there are other ways to construct a fourth order quadrupole invariant, e.g., $[\alpha \times \alpha]_2 \cdot [ \alpha \times \alpha]_2 $ and $[\alpha \times \alpha]_4 \cdot [ \alpha \times \alpha]_4$ but they are all proportional to $\beta^4$.

\subsubsection{Landau-like expansion}

Since the distribution $P(T,\alpha_{2\mu})$  is invariant under rotations,
its logarithm can be expanded in quadrupole invariants. In the spirit of Landau theory of shape transitions~\cite{Levit1984,Alhassid1986}, we carry out this expansion to fourth order using the invariants in Eqs.~\eqref{eq:invariants-intrinsic}\footnote{In the Landau theory developed in Refs.~\cite{Levit1984,Alhassid1986}, the Helmholtz free energy $F(T,\beta,\gamma)$ was expanded in the invariants to fourth order and the quadrupole shape fluctuations were described by the distribution $\propto \exp[-F(T,\beta,\gamma)/T]$. Thus $\ln P(T,\beta,\gamma)$ corresponds  to $-F(T,\beta,\gamma)/T$ up to an additive constant.}.  This leads  to the following probability distribution
\be\label{eq:landau-P}
 P(T, \beta, \gamma) = \mathcal{N}(T) e^{ -a(T) \beta^2 - b(T) \beta^3 \cos (3\gamma) - c(T) \beta^4 } \;,
\ee
where $a$, $b$, and $c$ are temperature-dependent parameters and $\mathcal N$ is a normalization constant.  The expectation value of a function $f(\beta,\gamma)$ that depends on the intrinsic deformation parameters $\beta,\gamma$ is given by 
 \be\label{landau-average}
  \langle f(\beta,\gamma) \rangle_L \equiv 4\pi^2 \int \der{\beta} \der{\gamma} \beta^4 |\sin(3\gamma)| f(\beta,\gamma) P(T,\beta,\gamma) \;,
\ee
where we have used the metric (\ref{metric}) and the subscript $L$ denotes an expectation value with respect to the distribution (\ref{eq:landau-P}) obtained in a Landau-like expansion.
In calculating the expectation values of the three quadrupole invariants in (\ref{eq:invariants-intrinsic}), the integration over $\gamma$ can be done analytically; see Eqs.~(\ref{gamma-int}) and (\ref{Cnm}) in Appendix~\ref{app:formulas}. The normalization constant $\mathcal N$ in (\ref{eq:landau-P}) is determined as a function of $a,b,c$ from the normalization condition $\langle 1 \rangle_L=1$.

The expansion parameters $a$, $b$, and $c$ in Eq.~(\ref{eq:landau-P}) are determined from the expectation values of the three quadrupole invariants. The latter can be calculated in AFMC using their relations to moments of the axial quadrupole operator $\hat Q_{20}$ in the laboratory frame 
\be
  \langle \hat Q_{20}^n \rangle = \int \der{q_{20}} q_{20}^n P(q_{20}) \;
\ee
as follows~\cite{Alhassid2014,Gilbreth2018}
\begin{subequations}\label{eq:invariants-lab}
\be
  \langle \hat Q \cdot \hat Q \rangle = 5 \langle \hat Q_{20}^2 \rangle,
\ee
\be
  \langle [\hat Q \times \hat Q]_2 \cdot \hat Q \rangle = -5\sqrt{\frac{7}{2}} \langle \hat Q_{20}^3 \rangle,
\ee
and
\be
  \langle (\hat Q \cdot \hat Q)^2 \rangle = \frac{35}{3} \langle \hat Q_{20}^4 \rangle \;.
\ee
\end{subequations}

Matching the quadrupole invariants computed using the distribution \eqref{eq:landau-P} with the invariants determined from the AFMC calculation using Eqs.~\eqref{eq:invariants-lab}, we obtain a set of nonlinear equations for $a,b,c$
\begin{subequations} \label{nleqns}
  \be
  \chi^2 \langle \beta^2 \rangle_L = 5 \langle \hat Q_{20}^2 \rangle \,,
  \ee
  \be
  \chi^3 \langle \beta^3 \cos(3\gamma)\rangle_L= \frac{35}{2} \langle \hat Q_{20}^3 \rangle \,,
  \ee
  \be
  \chi^4 \langle \beta^4 \rangle_L = \frac{35}{3}  \langle \hat Q_{20}^4 \rangle \,,
  \ee
\end{subequations}
where  $\chi= \frac{3}{\sqrt{5 \pi}} r_0^2 A^{5/3}$  [see Eq.~(\ref{q-to-alpha})].

\subsubsection{Validation of the Landau-like expansion}

In deriving the distribution (\ref{eq:landau-P}), we expanded the logarithm of $P(T,\beta,\gamma)$ in the quadrupole invariants to fourth order. In principle, higher-order invariants also contribute to this expansion. To test the validity of the fourth-order expansion, we can rewrite the distribution (\ref{eq:landau-P}) in terms of the lab-frame deformation variables $\alpha_{2\mu}$
 \be\label{eq:landau-lab}
  P(T, \alpha_{2\mu}) = \mathcal{N}(T) e^{ -a(T) \alpha\cdot\alpha + b(T)\sqrt{\frac{7}{2}} [\alpha\times\alpha]_2\cdot \alpha  - c(T) (\alpha\cdot\alpha)^2 } \;,
\ee
where we have used Eqs.~(\ref{eq:invariants-intrinsic}). 
We can then integrate over the four variables $\alpha_{2\mu}$ with $\mu\neq 0$ to determine the marginal distribution $P(T,\alpha_{20})$ and thus the distribution $P(q_{20})$ of the axial quadrupole $q_{20}$ in the laboratory frame. This distribution can be compared directly with the AFMC distribution $P(q_{20})$. 

\begin{figure}
  \begin{center}\includegraphics{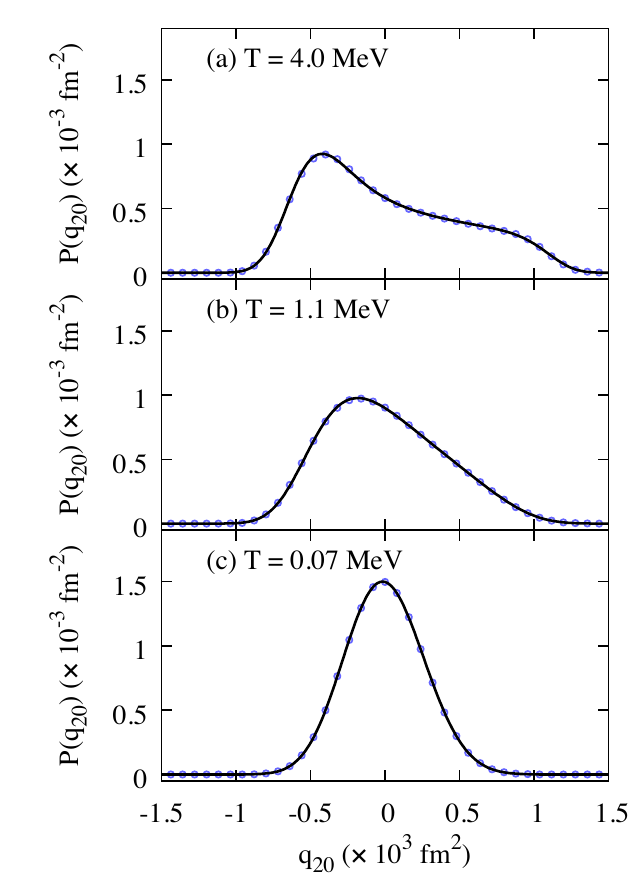}\end{center}
  \caption{The lab-frame axial quadrupole distribution $P(q_{20})$ for $^{154}$Sm at three distinct temperatures: (a) a high temperature $T=4$ MeV, (b) an intermediate temperature $T=1.1$ MeV near the shape transition, and (c) a low temperature $T=0.07$ MeV.  Solid lines are the marginal distributions $P(q_{20})$ obtained from the Landau-like expansion of the intrinsic shape distribution [Eq.~(\ref{eq:landau-P})], with parameters $a,b,c$ determined from the AFMC moments of $q_{2,0}$.  Open circles are the direct AFMC calculation of $P(q_{2,0})$ using Eqs.~(\ref{q-dist-AFMC}) and (\ref{Fourier}).  For clarity, only every fifth AFMC point has been included in the plot. The uncertainties in the AFMC results are smaller than the size of the symbols.}
  \label{fig:pqdist}
\end{figure}

In Fig.~\ref{fig:pqdist} we compare the distribution $P(q_{20})$ calculated from the marginal distribution of  Eq.~(\ref{eq:landau-lab}) (solid line) with the corresponding AFMC distribution (open circles) for $^{154}$Sm.
 At the resolution seen in the figure, the agreement is perfect. We conclude that the fourth-order Landau-like expansion is sufficient at all temperatures. 

\subsection{Applications to samarium isotopes}

We demonstrate our method for computing the intrinsic shape distribution $P(T,\beta,\gamma)$  for the family of even-mass samarium isotopes $^{148-154}$Sm, which are known to exhibit a crossover from spherical to deformed shapes.

Our single-particle shell-model space includes the orbitals $0\textrm{g}_{7/2}$, $1\textrm{d}_{5/2}$, $1\textrm{d}_{3/2}$, $2\textrm{s}_{1/2}$, $0\textrm{h}_{11/2}$, and $1\textrm{f}_{7/2}$ for protons, and the orbitals $0\textrm{h}_{11/2}$, $0\textrm{h}_{9/2}$, $1\textrm{f}_{7/2}$, $1\textrm{f}_{5/2}$, $2\textrm{p}_{3/2}$, $2\textrm{p}_{1/2}$, $0\textrm{i}_{13/2}$, and $1\textrm{g}_{9/2}$ for neutrons.
The single-particle energies and wave functions were obtained from a Woods-Saxon potential plus a spin-orbit interaction using the parameters of Ref.~\cite{Bohr1975}.
The interaction is a multipole-multipole interaction obtained by expanding a separable surface-peaked interaction up to the hexadecupole term, plus a monopole pairing interaction using the coupling parameters given in Ref.~\cite{Ozen2013}.

We estimate the statistical errors in our AFMC results using the block jackknife method (the method is described briefly in Appendix~\ref{jackknife}).  At each temperature $T$, we use an imaginary-time slice of $\Delta\beta=1/64$ MeV$^{-1}$ and $5120$ Monte Carlo samples, consisting of $128$ independent Monte Carlo walkers (on different CPUs), each composed of $40$ samples taken after thermalization. We chose a sufficiently large number of decorrelation sweeps  for the samples to be generally decorrelated. However, we observed that for the more deformed isotopes, decorrelation of the moments $\langle \hat Q_{20}^n \rangle$ was difficult to achieve. To obtain the correct uncertainty estimates, we chose in our jackknife method each independent 40-sample walker as a block over which we averaged all observables used in the next steps of the analysis.

\subsubsection{Moments of $\hat Q_{20}$ and the expansion parameters $a,b,c$}

\begin{figure}                     
  \begin{center}\includegraphics{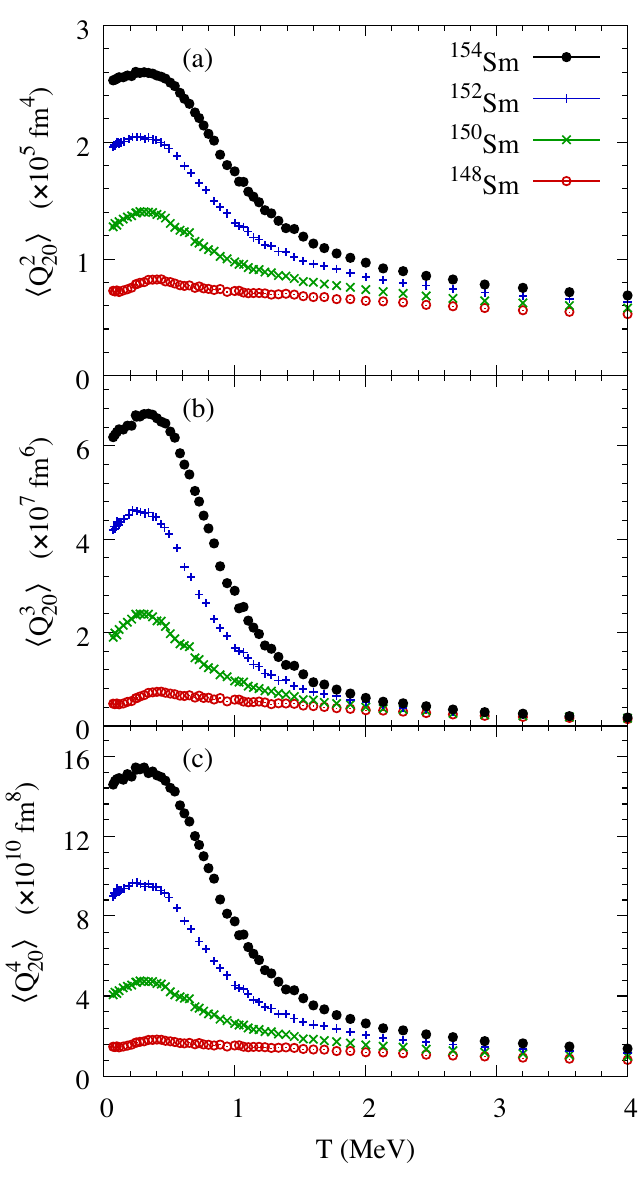}\end{center}
  \caption{The second [panel (a)], third [panel (b)] and fourth [panel (c)] moments of $\hat Q_{20}$, evaluated from the AFMC distributions $P(q_{20})$  as function of temperature $T$ for the even-mass samarium isotopes $^{148-154}$Sm.}
  \label{fig:qmoments}
\end{figure}

\begin{figure}
  \begin{center}\includegraphics{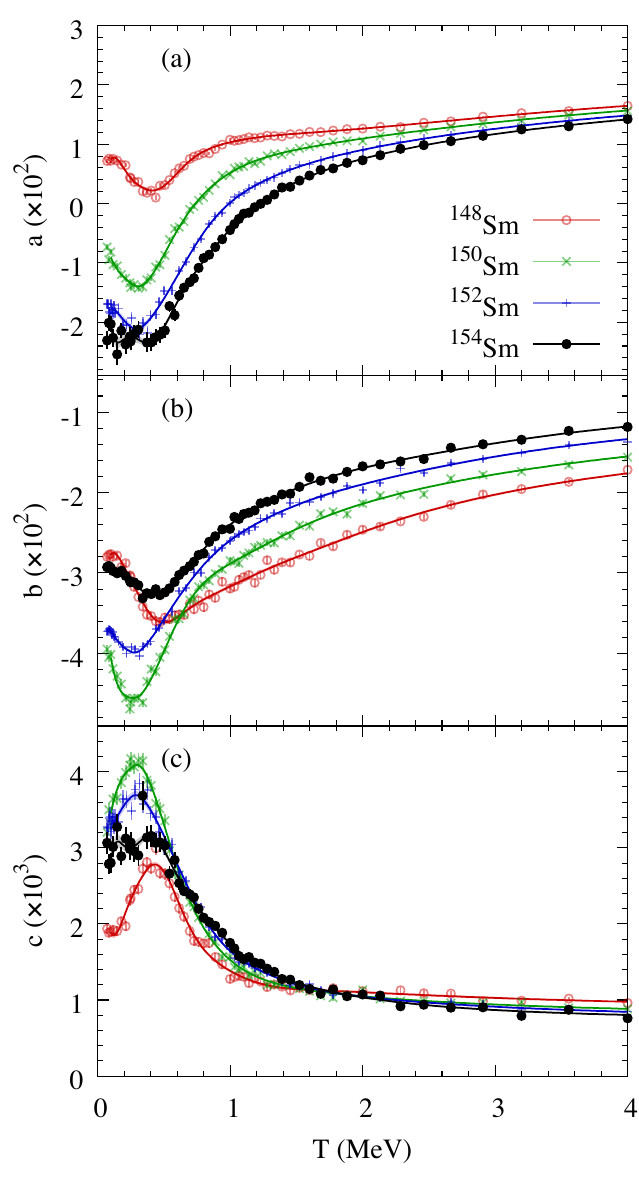}\end{center}
  \caption{The expansion parameters $a$ [panel (a)], $b$ [panel (b)], and $c$ [panel (c)] vs.~temperature $T$ for the even-mass samarium isotopes $^{148-154}$Sm, as determined from the moments in Fig.~\ref{fig:qmoments} by solving Eqs.~(\ref{nleqns}) (open circles). The solid lines describe the smoothing spline interpolation (see text).}
  \label{fig:abc}
\end{figure}

The second, third and fourth moments of $\hat Q_{20}$ evaluated from the AFMC distribution $P(q_{20})$ are shown in Fig.~\ref{fig:qmoments} as a function of temperature. In these results, we scaled $\hat Q_{20}$ by a factor of $2$ to account for core polarization effects.  At any given temperature $T$, the moments increase with the number of neutrons.

We determined the parameters $a$, $b$, and $c$ by solving Eqs.~\eqref{nleqns} to match the quadrupole invariants computed using the distribution \eqref{eq:landau-P} with the AFMC moments $\langle \hat Q_{20}^n \rangle$ calculated from $P(q_{20})$ for $n=2,3,4$. Figure~\ref{fig:abc} shows the  expansion coefficients $a,b,c$ as a function of temperature for the four even-mass samarium isotopes $^{148-154}$Sm.

\subsubsection{Intrinsic quadrupole shape distributions at fixed temperature}

\begin{figure*}[bth]
  \begin{center}\includegraphics{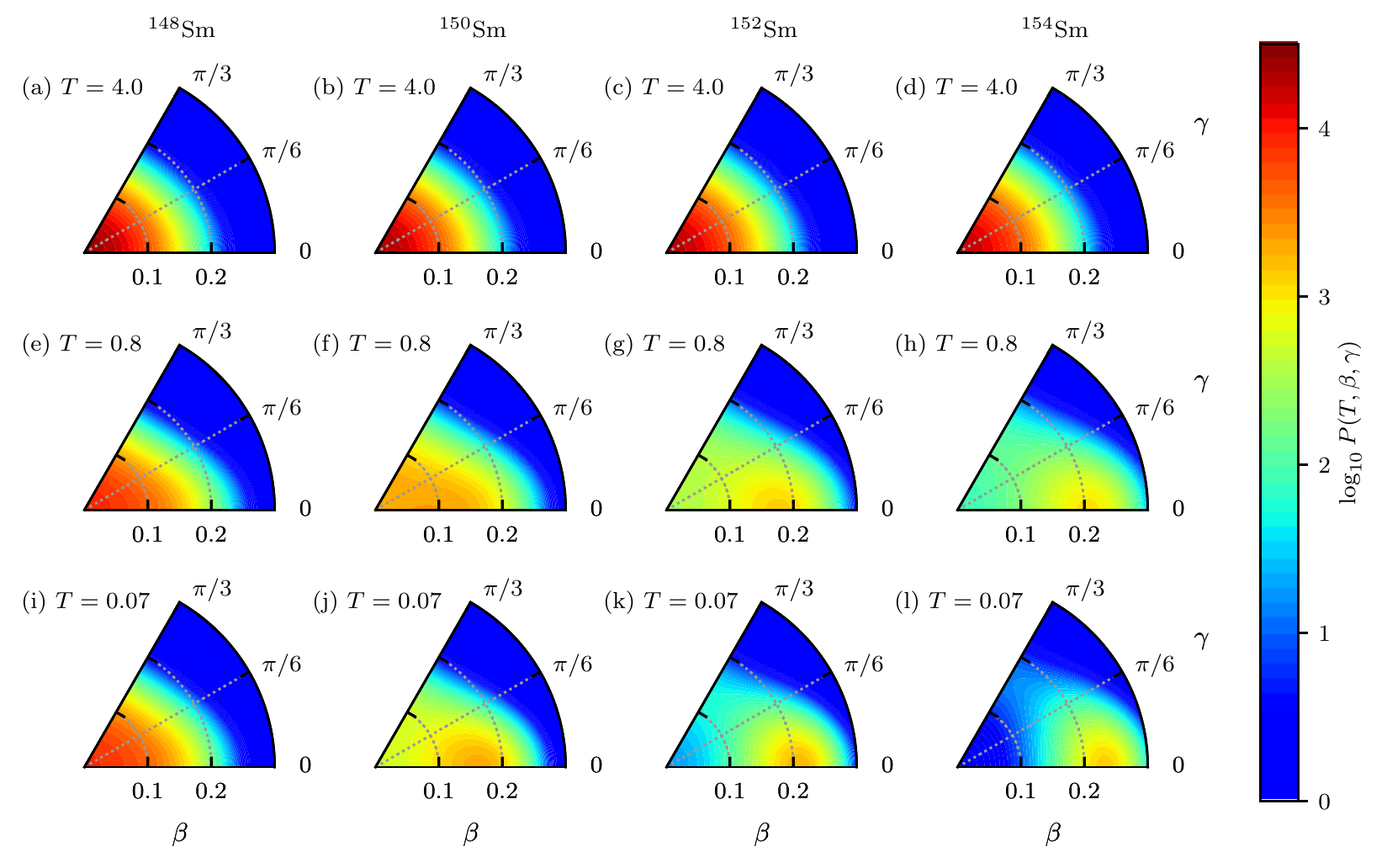}\end{center}
  \caption{Distributions $P(T,\beta,\gamma)$ (shown in a logarithmic scale) in the $\beta-\gamma$ plane for the even-mass samarium isotopes at different temperatures: a high temperature $T=4$ MeV [panels (a)-(d)], an intermediate temperature $T=0.8$ MeV [panels (e)-(h)], and a low temperature $T=0.07$ MeV [panels (i)-(l)]. A thermal shape transition from prolate to spherical shape is evident for all but the spherical nucleus $^{148}$Sm as the temperature increases. A quantum shape transition from a spherical to a prolate shape is also observed near the ground state ($T=0.07$~MeV) as the neutron number increases.}
  \label{fig:prob_distribution}
\end{figure*}

In Fig.~\ref{fig:prob_distribution} we show  $\log_{10} P(T,\beta,\gamma)$ in the $\beta-\gamma$ plane for the four even-mass samarium isotopes $^{148-154}$Sm at a low temperature $T=0.07$ MeV, an intermediate temperature $T=0.8$ MeV and a high temperature $T=4$ MeV.  The maxima of these distributions mimic the shape transitions that are usually observed in a mean-field approximation but within CI shell model approach that takes into account correlations in full. 
The signature of a thermal shape transition from prolate to spherical as the temperature increases is clearly seen in $^{152,154}$Sm which are dominated by a prolate deformation in their ground state.  In contrast, no thermal shape transition is observed in the spherical nucleus $^{148}$Sm.
The transitional nucleus $^{150}$Sm undergoes a thermal shape transition, although it is not as distinctive as for the heavier samarium isotopes. 
We also observe a quantum shape transition of the ground state (described here by the low-temperature distributions at $T=0.07$~MeV) from a spherical shape to a prolate shape as we increase the number of neutrons between $^{148}$Sm and $^{154}$Sm.

\begin{figure*}[bth]
  \begin{center}\includegraphics{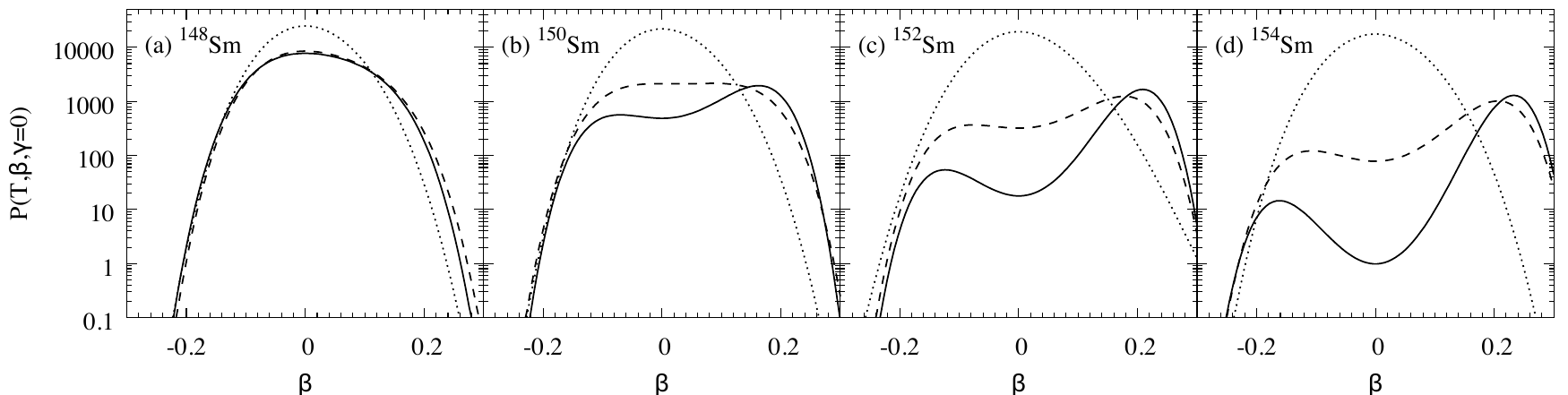}\end{center}
  \caption{The distribution $P(T,\beta,\gamma=0)$ (shown on a logarithmic scale) as a function of the axial deformation parameter $\beta$ for the even-mass samarium isotopes (a) $^{148}$Sm, (b) $^{150}$Sm, (c) $^{152}$Sm, and (d) $^{154}$Sm.
 The solid, dashed and doted lines correspond, respectively, to  temperatures of $T=0.07$~MeV, $T=0.8$~MeV, and  $T=4$~MeV.}
  \label{fig:pq_beta}
\end{figure*}

In Fig.~\ref{fig:pq_beta} we show on a logarithmic scale the distributions $P(T,\beta,\gamma=0)$ as a function of the axial deformation $\beta$ (negative values of $\beta$ describe axial deformations with $\gamma=\pi/3$) for $^{148-154}$Sm at the same temperatures as in Fig.~\ref{fig:prob_distribution}.  Following the maxima of these distributions, we again observe that $^{148}$Sm is spherical at all temperatures while $^{152,154}$Sm exhibit a clear shape transition from a prolate to a spherical shape as the temperature increases.  The transitional nucleus $^{150}$Sm also undergoes a thermal shape transition but the shape distribution at the intermediate temperature is rather flat for a wide range of $\beta$ values, reflecting coexistence of shapes. 

\begin{figure}[bth]
  \begin{center}\includegraphics{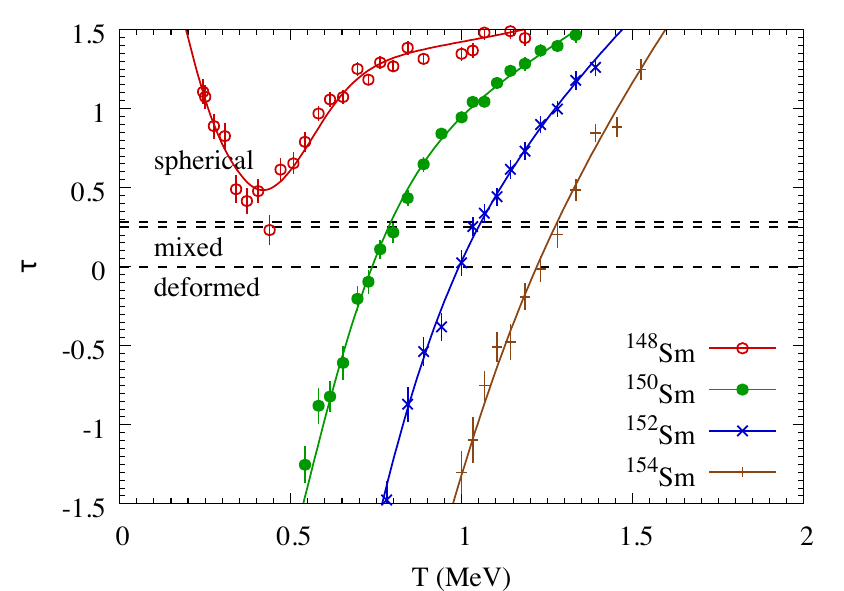}\end{center}
  \caption{The parameter $\tau=ac/b^2$ as a function of temperature $T$ for the even-mass samarium isotopes $^{148-154}$Sm. The values computed from the spline interpolation of $a,b$, and $c$ are described by the solid lines.  $\tau <0$  describes deformed shapes while $\tau>9/32$ describes spherical shapes.   The interval $0 < \tau <9/32$ is a mixed regime with a first-order shape transition at $\tau=1/4$.}
  \label{fig:tau}
\end{figure}

The topography of the distribution $P(T,\beta,\gamma)$ of Eq.~(\ref{eq:landau-P})  is completely determined by the dimensionless parameter $\tau = ac/b^2$~\cite{Levit1984,Alhassid1986}.\footnote{The stationary points of the distribution (\ref{eq:landau-P}) are axial with $\gamma=0$ ($\beta>0$) or $\gamma=\pi/3$ ($\beta <0$), and hence can be characterized (up to an overall scale)  by a single parameter $\tau$.}  In Fig.~\ref{fig:tau} we show $\tau$ as a function of temperature $T$ for the four even-mass samarium isotopes $^{148-154}$Sm.
 In the Landau theory of quadrupole shape transitions the spherical and prolate maxima of $P(T,\beta,\gamma)$ coexist as local maxima within the interval  $\tau= [0, 9/32]$ (shown as the ``mixed" region in the figure) with a first-order shape transition between the spherical and prolate shapes occurring at $\tau=1/4$.  According to our AFMC calculations, these shape transitions in $^{150}$Sm, $^{152}$Sm and  $^{154}$Sm occur, respectively, at temperatures of $T = 0.81$~MeV, $T = 1.03$~MeV, and $T = 1.29$~MeV. The corresponding transition temperatures according to the HFB calculations of Ref.~\cite{Gilbreth2018} are $T=0.74$~MeV, $T=0.94$~MeV, and $T=1.10$~MeV, respectively. It is interesting to note that $^{148}$Sm almost undergoes a shape transition as the temperature decreases to just below $0.5$~MeV.  However, as the temperature continues to decrease, $\tau$ increases again since the pairing interaction, which dominates at low temperatures in $^{148}$Sm, favors a spherical shape. 

\begin{figure}[bth]
  \begin{center}\includegraphics{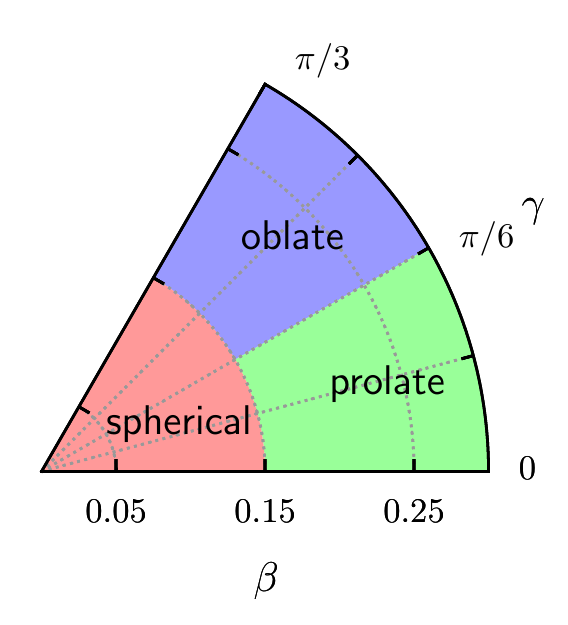}\end{center}
  \caption{Partition of the $(\beta, \gamma)$ plane into spherical, prolate, and oblate regions.}
  \label{fig:shapes}
\end{figure}

\begin{figure*}[bth]
  \begin{center}\includegraphics{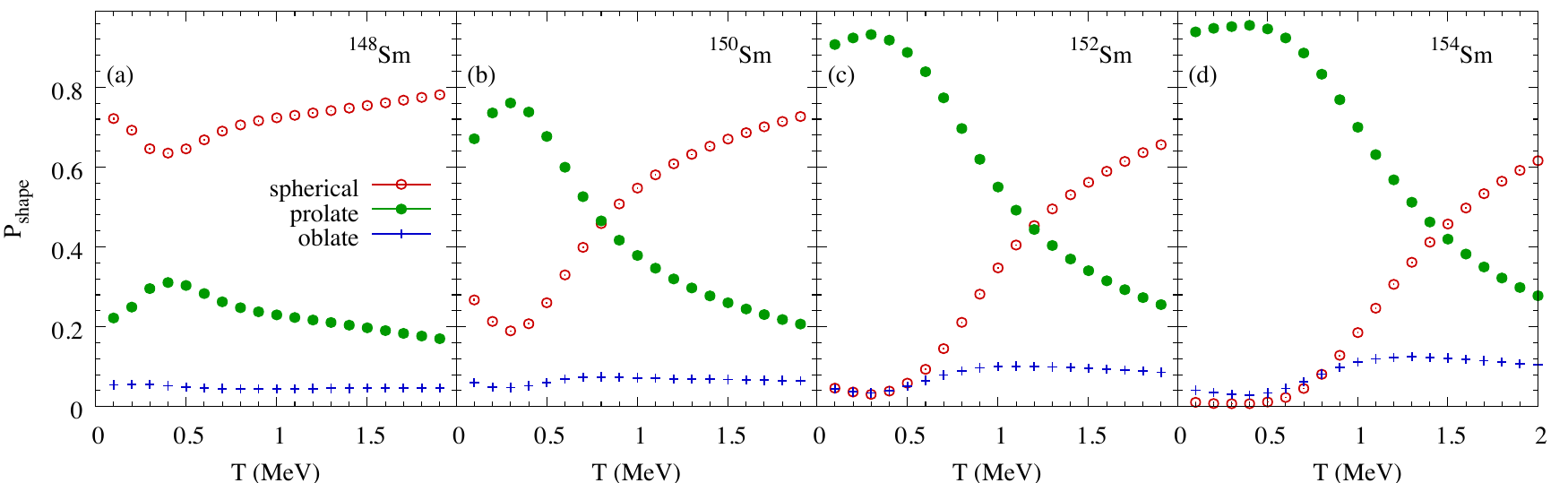}\end{center}
  \caption{The probabilities of spherical (open circles), prolate (solid circles) and oblate (pluses) shapes (defined as in Fig.~\ref{fig:shapes}) as a function of temperature $T$ for the even-mass samarium isotopes (a) $^{148}$Sm, (b) $^{150}$Sm, (c) $^{152}$Sm, and (d) $^{154}$Sm.}
  \label{fig:prob_shapes}
\end{figure*}

To facilitate the presentation of our results, we divide the $(\beta, \gamma)$-plane into three distinct regions, which represent spherical, prolate, and oblate shapes as in Fig.~\ref{fig:shapes} with $\beta_0=0.15$ separating between the spherical and deformed regions.
The probability of each of the three regions is determined by integrating the probability density $P(T,\beta,\gamma)$ with the corresponding metric over each of the regions 
\be\label{eq:integrated-P}
  P_\textrm{shape}(T) = 4\pi^2 \int_\textrm{shape} \!\!\!\!\!\!\!\! \der\beta \der\gamma \beta^4 |\sin(3\gamma)| P(T, \beta, \gamma) \;.
\ee
Here ``shape'' refers to any of the three regions --- spherical, prolate, or oblate --- as defined in Fig.~\ref{fig:shapes}.
The sum of these three shape probabilities is equal to 1.

The integrals over the intrinsic deformation coordinates $\beta, \gamma$ were approximated using the compound trapezoidal rule on a $20 \times 20$ mesh extending up to $\beta_\mathrm{max} = 0.3$.
An exception to this were the integrals in $\langle \beta^m \cos^n(3\gamma)\rangle_L$, for which the integration over $\gamma$ is done analytically; see Eqs.~(\ref{gamma-int}) and (\ref{Cnm}) in Appendix \ref{app:formulas}. This number of mesh points and the cutoff $\beta_\mathrm{max}$ were determined by requiring convergence of the integrals for the samarium isotopes; other nuclei may require a larger number of mesh points and/or a larger cutoff $\beta_\mathrm{max}$.

In Fig.~\ref{fig:prob_shapes} we show the spherical (open circles), prolate (solid circles) and oblate (pluses) shape probabilities as a function of temperature $T$ for the four samarium isotopes $^{148-154}$Sm.  In the isotopes that are deformed in their ground state ($^{150-154}$Sm) we observe a competition between the prolate and spherical shapes. Prolate shapes dominate at low temperatures and spherical shapes at higher temperatures. The prolate and spherical shape probabilities cross at a temperature that is higher for the heavier isotopes which are more strongly deformed in their ground state. In $^{148}$Sm, the spherical region dominates at all temperatures but its probability has a minimum at a temperature of $T \sim 0.4$ MeV that is close to the temperature where the parameter $\tau$ has a minimum (see Fig.~\ref{fig:tau}).  The contribution from oblate shapes is small for all four isotopes.  In the most deformed isotope $^{154}$Sm, it slightly exceeds the spherical probability at low temperatures.

\section{State densities versus intrinsic deformation}\label{sec:state_densities}

In this section we discuss the calculation of the state density as a function of intrinsic deformation $\beta,\gamma$ and excitation energy $E_x$ from the intrinsic shape distribution $P(T,\beta,\gamma)$.  

\subsection{Saddle-point approximation}

The state density $ \rho(E,\beta,\gamma)$ at energy $E$ and given intrinsic deformation parameters $\beta,\gamma$ is given by the inverse Laplace transform of the shape-dependent partition function $Z(T,\beta,\gamma)$ 
\be\label{Laplace}
  \rho(E, \beta,\gamma) = \frac{1}{2 \pi i} \int_{-i\infty}^{i\infty} d (1/T) \; e^{E/T} Z(T,\beta,\gamma) \;.
\ee
We calculate the shape-dependent partition function from the distribution $P(T,\beta,\gamma)$ using the relation
\be\label{shape-partition}
P(T,\beta,\gamma)={ Z(T,\beta,\gamma)\over Z(T)} \;,
\ee
where $Z(T)$ is the total partition function calculated from the thermal energy $E(T)$ as in Ref.~\cite{Nakada1997}.

To determine the average state density at a given deformation, we evaluate the integral in (\ref{Laplace}) using the saddle-point approximation 
\be\label{eq:rhoq}
 \rho(E, \beta, \gamma)
 \approx \frac{e^{S(T, \beta, \gamma)}}{\sqrt{2\pi T^2 C(T, \beta, \gamma)}} \;.
\ee
Here 
\be\label{entropy}
  S(T,\beta,\gamma) = \ln Z(T,\beta,\gamma)  + E/T
\ee
and
\be\label{heat-capacity}
C(T,\beta,\gamma) = T \frac{\partial S(T,\beta,\gamma)}{\partial T} 
\ee
are, respectively, the entropy and heat capacity at the corresponding deformation $\beta,\gamma$. The temperature $T$ in (\ref{entropy})  and (\ref{heat-capacity}) is determined as a function of energy $E$ and deformation $\beta,\gamma$ from the saddle-point condition 
\be\label{eq:saddlepoint}
  E(T,\beta,\gamma) \equiv T^2 \frac{\partial \ln Z(T,\beta,\gamma)}{\partial T} = E \;.
\ee
Substituting $T=T(E,\beta,\gamma)$ in (\ref{entropy}) and (\ref{heat-capacity}), we determine the state density in (\ref{eq:rhoq}) as a function of $E,\beta,\gamma$.  The corresponding excitation energy is calculated from $E_x=E-E_0$, where $E_0$ is the ground-state energy. 

The shape-dependent partition function $Z(T,\beta,\gamma)$ depends on the expansion coefficients $a,b,c$ through Eqs.~(\ref{shape-partition}) and (\ref{eq:landau-P}). Consequently, the shape-dependent entropy in (\ref{entropy}) depends on the first derivatives  $da/dT, db/dT$ and $dc/dT$, while the heat capacity in (\ref{heat-capacity}) depends on both the first derivatives and the second derivatives $d^2a/dT^2, d^2b/dT^2$ and $d^2c/dT^2$. The explicit expressions are given in Appendix~\ref{app:formulas}. 

In analogy with Eq.~\eqref{eq:integrated-P}, we can define state densities that correspond to each of the three deformation regions in Fig.~\ref{fig:shapes} by integrating the deformation-dependent state density over the corresponding regions
\be\label{eq:integrated-rho}
  \rho_\textrm{shape}(E) = 4\pi^2 \int_\textrm{shape} \!\!\!\!\!\!\!\! \der\beta \der\gamma \beta^4 |\sin(3\gamma)| \rho(E, \beta, \gamma) \;.
\ee

\subsection{Application to samarium isotopes} 

\subsubsection{Spline fits for $a,b,c$ and their temperature derivatives}

The coefficients $a$, $b$, and $c$ which characterize the probability distribution \eqref{eq:landau-P} and which are determined from the AFMC moments of $q_{20}$ in the laboratory frame, have statistical errors that are significantly amplified when taking the first derivatives and especially their second derivatives with respect to temperature. These derivatives are required in the calculation of the shape-dependent energy, entropy and heat capacity in Eqs.~(\ref{eq:saddlepoint}), (\ref{entropy}), and (\ref{heat-capacity}). To reduce the uncertainties in the derivatives of $a,b,c$, we fit a cubic smoothing spline for each of the coefficients, and use this spline for both interpolation between the sampled temperature values and for the derivatives. 

\begin{figure*}[bth]
  \begin{center}\includegraphics{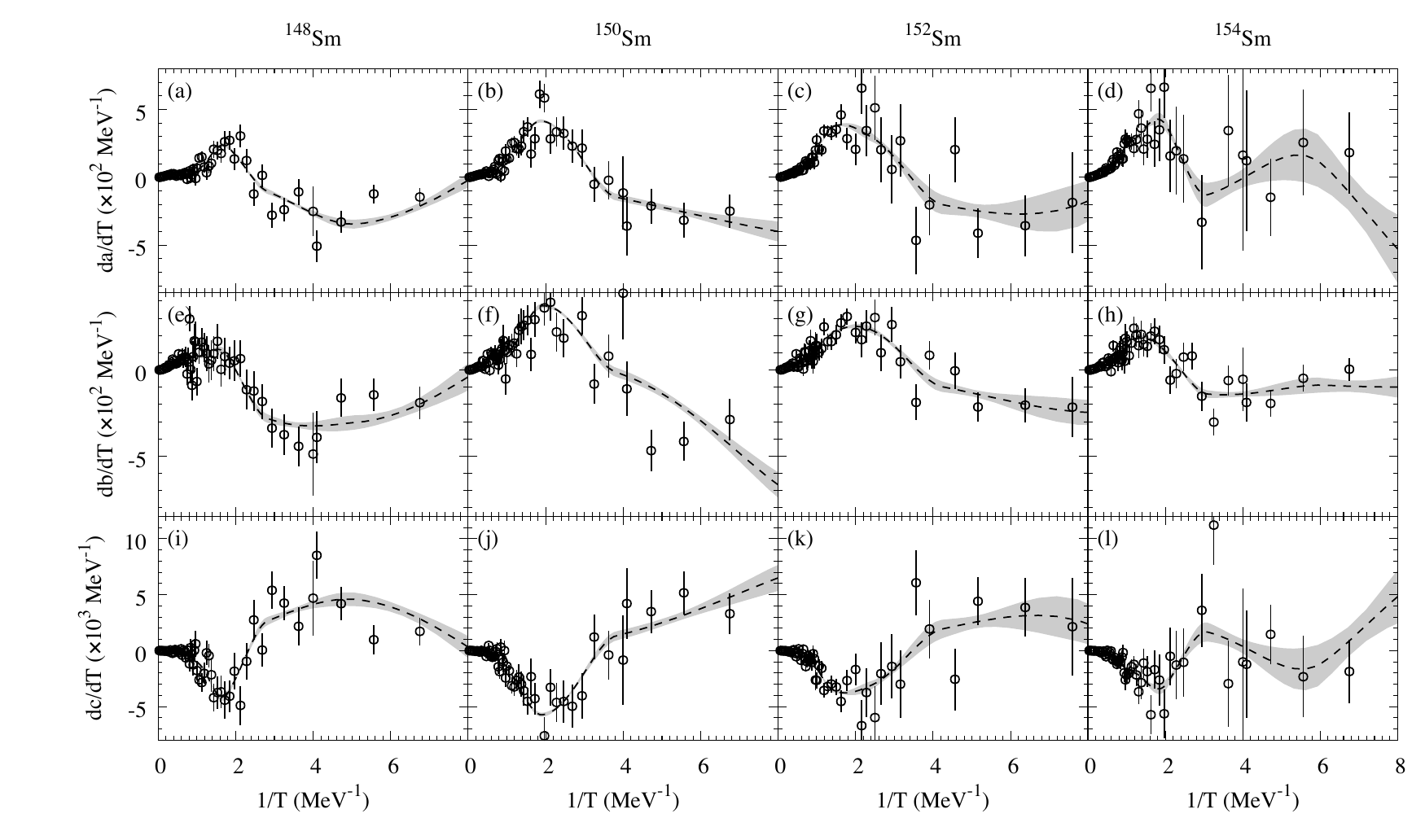}\end{center}
  \caption{The first derivatives $da/dT$ [panels (a)-(d)], $db/dT$ [panels (e)-(h)], and $dc/dT$ [panels (i)-(l)] of the Landau-like expansion parameters calculated in AFMC (open circles with error bars) and the derivatives of the smoothing spline interpolation of $a,b,c$ (dashed lines with uncertainties shown as shaded bands).}
  \label{fig:abc_derivatives}
\end{figure*}

The least-squares spline fit is made for  each of the coefficients $a$, $b$, and $c$ as function of $1/T$.
The number of knot points for the spline is chosen so that the reduced $\chi^2$ of the fit for each coefficient is between $1$ and $1.5$. In our computations, this translated to seven ($^{150}$Sm), ten ($^{148,152}$Sm), or eleven ($^{154}$Sm) spline segments. The knot points are placed so that the points extracted from the moments are partitioned between the spline intervals as evenly as possible. We set natural boundary conditions for the spline, i.e.~the second derivative is required to vanish at both ends. The cubic spline fits are shown by the solid lines in Fig.~\ref{fig:abc}.

In Fig.~\ref{fig:abc_derivatives}, we show the derivatives $da/dT,db/dT$ and $dc/dT$ as a function of $1/T$ obtained from the fitted splines (dashed lines with shaded bands describing the statistical error) and compared to the derivatives calculated by direct numerical differentiation (open circles with statistical errors). 

\subsubsection{Shape-dependent state densities} 

Using Eq.~(\ref{shape-partition}), the shape-dependent energy $E(T,\beta,\gamma)$ in the saddle-point condition Eq.~\eqref{eq:saddlepoint}, and the shape-dependent entropy $S(T,\beta,\gamma)$ and heat capacity $C(T,\beta,\gamma)$ in Eqs.~(\ref{entropy}) and (\ref{heat-capacity}) can be written as

\be\label{eq:Eq}
  E(T, \beta, \gamma)
  = E(T) + T^2 \frac{\partial}{\partial T} \ln P(T,\beta,\gamma) \;,
\ee
\be\label{eq:Sq}
  S(T, \beta, \gamma)
  = S(T)+ \ln P(T,\beta,\gamma) + T \frac{\partial}{\partial T} \ln P(T,\beta,\gamma) \;,
\ee
and
\be\label{eq:Cq}
\begin{split}
  C(T, \beta, \gamma) =& C(T)+  2 T \frac{\partial}{\partial T} \ln P(T,\beta,\gamma) \\
  &+ T^2 \frac{\partial^2}{\partial T^2} \ln P(T,\beta,\gamma) \;.
\end{split}
\ee

Here $E(T)$ is the total thermal energy calculated in AFMC from $\langle \hat H\rangle$,  $S(T) =  \ln Z(T) + E(T)/T$ is the canonical entropy and $C(T)$ is the canonical heat capacity $C(T)=dE/dT$.  To reduce the AFMC uncertainty of $C(T)$, we employed the method introduced in Ref.~\cite{Liu2001},  in which the same auxiliary-field configurations are used at inverse temperatures $\beta \pm \delta\beta$ to compute the numerical derivative of the total energy (taking into account correlated errors).

\begin{figure}[bth]
  \begin{center}\includegraphics{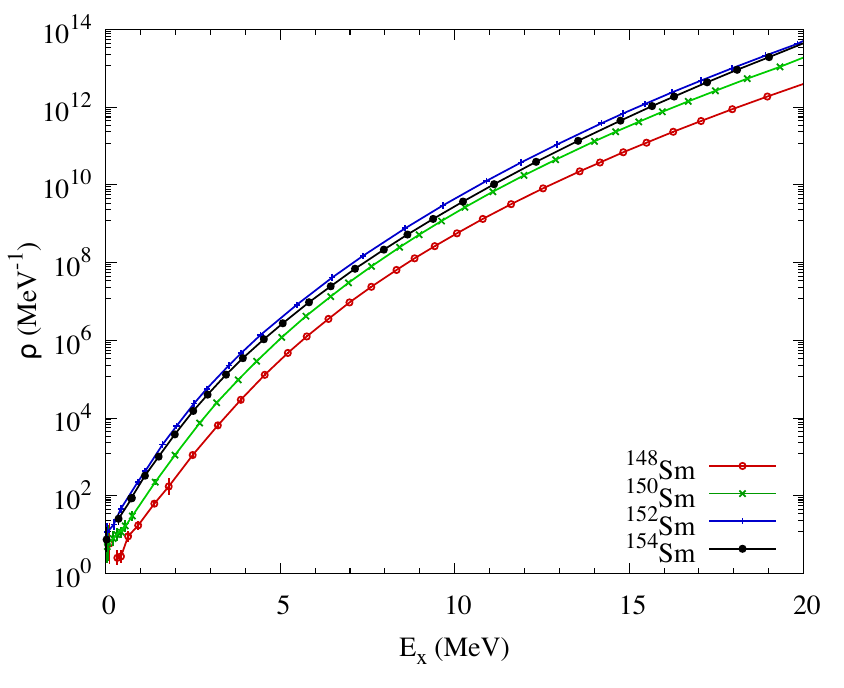}\end{center}
 \caption{The total state densities $\rho(E_x)$  computed directly from the thermal energy $E(T)$ for the even-mass samarium isotopes $^{148-154}$Sm.}
 \label{fig:state_density_total}
\end{figure}

Figure~\ref{fig:state_density_total} shows the total state densities $\rho(E_x)$ as a function of excitation energy $E_x$ for the four samarium isotopes, calculated directly from the thermal energy $E(T)$ as in Ref.~\cite{Nakada1997}. 

\begin{figure*}[bth]
  \begin{center}\includegraphics{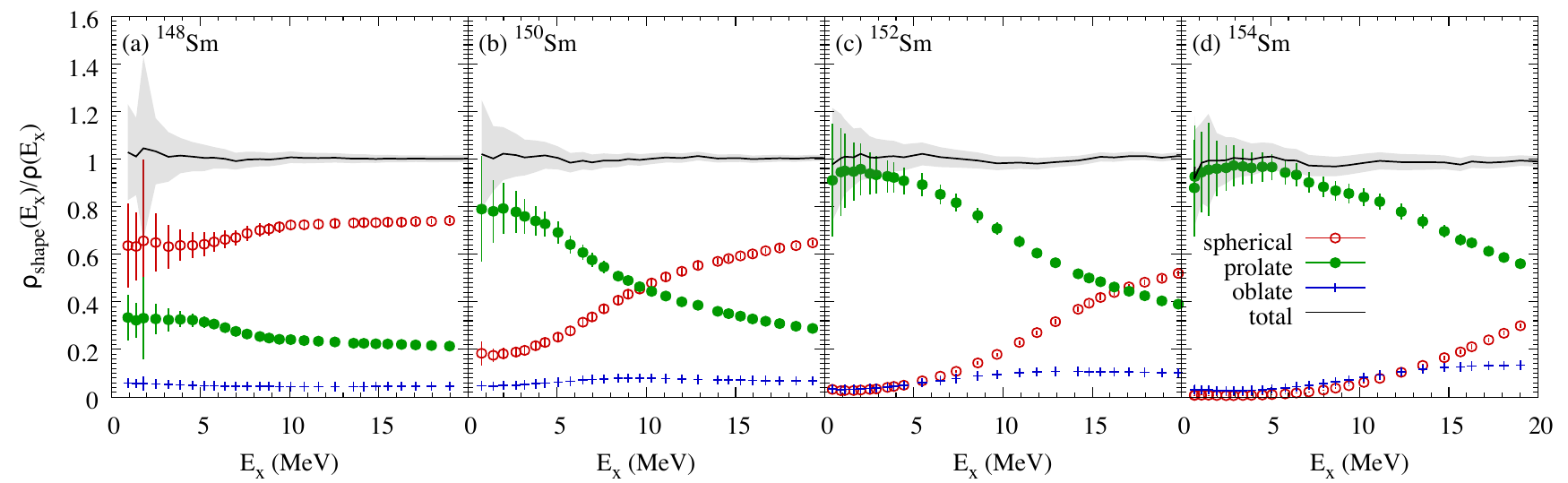}\end{center}
  \caption{The shape probabilities $ \rho_{\rm shape}(E_x)/\rho(E_x)$ as a function of excitation energy $E_x$ for each of the three regions in Fig.~\ref{fig:shapes}: spherical (open circles), prolate (solid circles) and oblate (es) for (a) $^{148}$Sm, (b) $^{150}$Sm, (c) $^{152}$Sm, and (d) $^{154}$Sm. The solid lines are the sum of these three probabilities with statistical errors shown by the shaded bands. The sum rule (\ref{sum_rule}) is satisfied within the statistical errors.}
  \label{fig:state_density_shapes}
\end{figure*}

Figure~\ref{fig:state_density_shapes} presents the main result of this work, showing (for the four even-mass samarium isotopes $^{148-154}$Sm) the ratios $\rho_{\rm shape}(E_x)/\rho(E_x)$ of the shape-dependent state densities $\rho_{\rm shape}(E_x)$ in Eq.~(\ref{eq:integrated-rho}) to the total state density $\rho(E_x)$ vs.~excitation energy $E_x$ for each of the three deformation regions of Fig.~\ref{fig:shapes} (i.e., spherical, prolate and oblate).  In the isotopes that are deformed in their ground state $^{150,152,154}$Sm, the prolate state density dominates at low excitation energies but the spherical state density exceeds it at above a certain excitation energy that becomes higher for the heavier isotopes.\footnote{We  note that the exact excitation energy for which the crossing of the spherical and prolate densities occur depends on the value of $\beta_0$ used to differentiate between the spherical and deformed regions in Fig.~\ref{fig:shapes}.}
 In the well-deformed nuclei  $^{152,154}$Sm the probability of the prolate shape is close to $1$ up to excitations of $E_x\sim 5$ MeV, while in the transitional nucleus $^{150}$Sm it is only $\sim 0.8$ up to $E_x \sim 3$ MeV. In the spherical nucleus $^{148}$Sm, the spherical state density dominates at all excitation energies although the prolate shape region makes a significant contribution. The contribution of the oblate shape is relatively small in all four isotopes. 

\subsubsection{Sum rule} 

 Integrating the shape-projected state density over all shapes $\beta,\gamma$ in the intrinsic frame should yield the total state density and can thus be compared with the total state density $\rho(E_x)$ of Fig.~\ref{fig:state_density_total}. Alternatively, the sum of the three shape probabilities (spherical, prolate and oblate regions in Fig.~\ref{fig:shapes}) should satisfy the sum rule 
 \be\label{sum_rule}
 \sum_{\rm shapes} \rho_{\rm shape}(E_x)/\rho(E_x) =1 \;.
 \ee
 These sums are shown for the four samarium isotopes by the solid lines in the figure with error bars indicated by the shaded gray bands. We find that the sum rule (\ref{sum_rule}) is satisfied within the error bars in all four isotopes. We note  that since the saddle-point approximation is used separately for each deformation $\beta,\gamma$,  the sum rule is not expected to be satisfied exactly and provides a non-trivial test of the accuracy of our method.

\section{Conclusion}\label{sec:conclusions}

We have presented a method for computing the nuclear state density as a function of the intrinsic quadrupole deformation and excitation energy
that preserves the rotational invariance of the Hamiltonian.  Specifically, the AFMC method is applied in the framework of the CI shell model to compute
the distribution of the axial mass quadrupole in the laboratory frame [defined by Eq.~(\ref{q-dist})]  which is then used to extract the intrinsic properties.

 In broader terms, this article describes a method to calculate energy-dependent statistical properties 
of a finite-size many-particle system that undergoes a symmetry-breaking phase transition in the thermodynamic limit. This phase transition is described by order parameters which in the low-temperature phase break a certain symmetry of the Hamiltonian. The challenge is to calculate the thermal distribution of the order parameters within a framework that preserves the exact symmetry and without invoking a mean-field approximation.  In the following, we assume that the order parameters are described by one-body operators that transform according to an irreducible representation of the corresponding symmetry group.  The important ingredients of the method are:

\noindent a) Construction of the marginal distribution with respect to one or more components of the order parameter by using a projection on the corresponding one-body operator.  

\noindent b) Determination of  the expectation values of low-order polynomial combinations of the order parameters that are invariant under the symmetry group. This is accomplished by relating these invariants to moments of the marginal distributions constructed in a).

 \noindent c) Expansion of the logarithm of the thermal distribution of the order parameters (i.e., the Helmholtz free energy) in the invariants described in b).  Such a Landau-like expansion is justified by the invariance of the this distribution under transformations of the symmetry group and is carried out up to the lowest order that is sufficient to describe the phase transition. The temperature-dependent parameters that appear in this expansion are determined from the expectation values of the invariants calculated in b). 
 
In the particular example discussed in this article, the symmetry group is the rotation group and the order parameters are the quadrupolar deformation tensor $q_{2\mu}=\chi \alpha_{2\mu}$ in the laboratory frame.  The marginal distribution is that of the axial quadrupole $q_{20}$ in the laboratory frame defined by Eq.~(\ref{q-dist}). This marginal distribution has been calculated using Eqs.~(\ref{q-dist-AFMC}) and (\ref{Fourier}) as described in Refs.~\cite{Alhassid2014,Gilbreth2018}.  We have used the AFMC computational scheme, but for smaller model spaces it could also have been done by standard matrix configuration-interaction methods.

We found remarkable simplifications in carrying out part b) for our application in that the marginal distribution $P(q_{20})$ of a single component of the quadrupolar tensor was sufficient to determine the expectation values of the three lowest order invariants [see Eqs.~(\ref{eq:invariants-lab})]. 
It is also remarkable that these three invariants turn out to be sufficient to construct a Landau-like expansion of $\ln P(T, \alpha_{2\mu})$ [see Eq.~(\ref{eq:landau-lab})] that describes the actual marginal distribution  $P(q_{20})$ to a very good accuracy (see Fig.~\ref{fig:pqdist}).

The example we studied in this article, the samarium isotope chain, is a paradigm  for the shape transition between spherical and deformed nuclei. As is known experimentally and supported by many studies using mean-field approximations, the lighter isotopes are spherical in their ground state and the heavier isotopes become progressively more deformed.  Besides confirming this behavior, our method describes how the deformation becomes progressively weaker at higher 
excitation energies.  In this respect, we confirm earlier studies showing that the transition from deformed to spherical shapes as the excitation energy increases is rather gradual and far from that characterized by a first-order phase transition predicted by pure mean-field theory.

\begin{acknowledgements}
 This work was supported in part by the U.S. DOE grant Nos.~DE-FG02-91ER40608 and DE-FG02-00ER411132. 
The research presented here used resources of the National Energy Research Scientific Computing Center, which is supported by the Office of Science of the U.S. Department of Energy under Contract No.~DE-AC02-05CH11231.  This work was also supported by the HPC facilities operated by, and the staff of, the Yale Center for Research Computing.\\
\end{acknowledgements}

\appendix

\section{Logarithmic derivatives of the shape-dependent probability $P(T,\beta,\gamma)$}\label{app:formulas}

The evaluation of the shape-dependent energy, entropy and heat capacity in Eqs.~(\ref{eq:Eq}), (\ref{eq:Sq}) and (\ref{eq:Cq}) require the first and second logarithmic derivatives of the distribution $P(T,\beta,\gamma)$ with respect to temperature. Here we express these derivatives in terms of derivatives of the Landau-like expansion coefficients $a,b,c$.

\begin{widetext}
\begin{equation}
  \frac{\partial}{\partial T} \ln P(T,\beta,\gamma)
  = \frac{\partial a}{\partial T} \bigl(\langle \beta^2 \rangle_L - \beta^2 \bigr)
  + \frac{\partial b}{\partial T} \bigl(\langle \beta^3 \cos (3\gamma) \rangle_L - \beta^3 \cos (3\gamma) \bigr)
  + \frac{\partial c}{\partial T} \bigl(\langle \beta^4 \rangle_L - \beta^4 \bigr)
\end{equation}
and
\begin{equation}
  \begin{split}
    \frac{\partial^2}{\partial T^2} \ln P(T,\beta,\gamma) =& \frac{\partial^2 a}{\partial T^2} \bigl(\langle \beta^2 \rangle_L - \beta^2 \bigr)
    + \frac{\partial^2 b}{\partial T^2} \bigl(\langle \beta^3 \cos (3\gamma) \rangle_L - \beta^3 \cos (3\gamma) \bigr)
    + \frac{\partial^2 c}{\partial T^2} \bigl(\langle \beta^4 \rangle_L - \beta^4 \bigr)\\
    &+ \biggl( \frac{\partial a}{\partial T} \biggr)^{\!\!2} \bigl(\langle \beta^2 \rangle_L^2 - \langle \beta^4 \rangle_L \bigr)
    + \biggl( \frac{\partial b}{\partial T} \biggr)^{\!\!2} \bigl(\langle \beta^3 \cos (3\gamma) \rangle_L^2 - \langle \beta^6 \cos^2 (3\gamma) \rangle_L \bigr)
    + \biggl( \frac{\partial c}{\partial T} \biggr)^{\!\!2} \bigl(\langle \beta^4 \rangle_L^2 - \langle \beta^8 \rangle_L \bigr) \\
    &+ 2 \frac{\partial a}{\partial T} \frac{\partial b}{\partial T}
    \bigl( \langle \beta^2 \rangle_L \langle \beta^3 \cos(3\gamma) \rangle_L - \langle \beta^5 \cos(3\gamma) \rangle_L \bigr)
    + 2 \frac{\partial a}{\partial T} \frac{\partial c}{\partial T}
    \bigl( \langle \beta^2 \rangle_L \langle \beta^4 \rangle_L - \langle \beta^6 \rangle_L \bigr)
    \\
    &+ 2 \frac{\partial b}{\partial T} \frac{\partial c}{\partial T}
    \bigl(\langle \beta^3 \cos(3\gamma) \rangle_L \langle \beta^4 \rangle_L - \langle \beta^7 \cos(3\gamma) \rangle_L \bigr),
  \end{split}
\end{equation}
where the expectation values $\langle \ldots\rangle_L$ are defined as in (\ref{landau-average}).
\end{widetext}

The integration over $\gamma$ in calculating the expectation values $ \langle \beta^m \cos^n(3\gamma) \rangle_L$ can be done analytically. This yields the formula
\begin{equation}\label{gamma-int}
  \langle \beta^m \cos^n(3\gamma) \rangle_L = \frac{\int_0^\infty \der\beta e^{-a\beta^2-c\beta^4} C_{nm}(\beta)}{\int_0^\infty \der\beta \beta e^{-a\beta^2-c\beta^4} \sinh (b\beta^3)},
\end{equation}
where the functions $C_{nm}(\beta)$ for $n=0,1,2$ are given by
\begin{subequations}\label{Cnm}
\begin{equation}
C_{0m} = \beta^{m+1} \sinh(b\beta^3)\;,
\end{equation}
\begin{equation}
C_{1m} = \frac{1}{b} \beta^{m-2} \sinh(b\beta^3) - \beta^{m+1} \cosh(b\beta^3)\;,
\end{equation}
and
\begin{equation}\begin{split}
C_{2m} = \beta^{m+1} \biggl[ &\biggl(1 + \frac{2}{b^2 \beta^6}\biggr) \sinh(b\beta^3) \\ &- \frac{2}{b \beta^3} \cosh(b\beta^3) \biggr].
\end{split}\end{equation}
\end{subequations}
The remaining quadratures over the axial deformation parameter $\beta$ are calculated numerically.

\section{The jackknife method}\label{jackknife}

The jackknife technique is a well-known method for variance and bias estimation in statistics.
Here we summarize the method, referring to Refs.~\cite{Young2014,Quenouille1949,Tukey1958} for more detail.

While the original motivation for the jackknife was to reduce the bias of statistical estimates, the procedure has an additional major advantage in case of complex computations.
It does not require computing analytical partial derivatives, in contrast to the traditional error propagation formula based on Taylor's expansion.
The jackknife method is particularly useful when the analytic form of the partial derivatives is intractable.

The jackknife method for estimating the uncertainty is straightforward.
Given a function $f(x, y, \cdots)$ and $N$ independent and identically distributed (i.i.d.) samples of its variables $(x, y, \cdots)$, one first leaves out the $i$-th sample $(x_i, y_i, \cdots)$ of the data (for each $i$ at a time), and computes the averages
\be
 (x^{(i)}, y^{(i)}, \cdots)
 = \bigg( \frac{1}{N-1} \sum_{j \ne i} x_j, \frac{1}{N-1} \sum_{j \ne i} y_j, \cdots \bigg)
\ee
for $i=1,\ldots,N$.
One then computes the function $f$ for each of these $N$ averages
\be
	f^{(i)} = f(x^{(i)}, y^{(i)}, \cdots) \;.
\ee
Finally, one uses the $N$ values $f^{(i)}$ to estimate the average value of the function
\begin{equation}\label{eq:jackknife-estimate}
 f_J = \frac{1}{N} \sum_i f^{(i)} \;,
\end{equation}
and its standard error
\begin{equation}\label{eq:jackknife-error}
 \delta f_J = \sqrt{\frac{N-1}{N} \sum_i \bigl( f_J - f^{(i)} \bigr)^2} \;.
\end{equation}
We note that (\ref{eq:jackknife-error}) differs from the usual error formula (which is used for uncorrelated values) by the factor $N-1$ because the values $f^{(i)}$, computed from averages of sets differing from one another only by one sample, are highly correlated.

The jackknife procedure is consistent with the standard error formula obtained by using the Taylor expansion for the function $f$ (see, e.g., in Ref.~\cite{Young2014}).
If the samples are correlated but can be divided into equally sized uncorrelated blocks of samples, the jackknife method can be applied to block averages of the variables $x, y,\ldots$ (this is the case where each block is generated by a Monte Carlo walk on a different CPU).  
It is easy to show that this is equivalent to leaving out consecutive non-overlapping blocks of samples (instead of single samples), also known as delete-$k$ jackknife.

\bibliographystyle{apsrev4-1}
%

\end{document}